# Emergence of field-induced memory effect in spin ices

Pramod K. Yadav[a*], Rajnikant Upadhyay[b], Rahul Kumar[c], Pavan Nukala[a] and Chandan Upadhyay[b]

[a]*Centre for Nano Science and Engineering, Indian Institute of Science, Bangalore-560012, India*
[b]*School of Materials Science and Technology, Indian Institute of Technology (Banaras Hindu University), Varanasi-221005, India*
[c]*School of Advanced Materials and Chemistry and Physics of Materials Unit, Jawaharlal Nehru Centre for Advanced Scientific Research, Bangalore-560064, India*

[*]Email: pramodyadav@iisc.ac.in; pyadav.rs.mst13@itbhu.ac.in

*Abstract:* Out-of-equilibrium investigation of strongly correlated materials deciphers the hidden equilibrium properties. Herein, we have investigated the out-of-equilibrium magnetic properties of polycrystalline $Dy_2Ti_2O_7$ and $Ho_2Ti_2O_7$ spin ices. The experimental results show the emergence of magnetic field-induced anomalous hysteresis observed only in temperature/magnetic field-dependent ac susceptibility measurements. The observed memory effect (anomalous thermomagnetic hysteresis) strongly depends on the driving thermal and non-thermal variables. Contrary, in the absence of the magnetic field, dipolar interaction induced Ising paramagnetic to spin ice crossover develops a liquid-gas transition type hysteresis below 4 K. Unlike field-induced hysteresis, it shows weak dependency on thermal and non-thermal variables. Due to the non-colinear spin structure, the applied dc bias magnetic field produces quench disorder sites in the cooperative Ising spin matrix and suppresses the spin-phonon coupling. These quench disorders create dynamic spin correlations governed by quantum fluctuations, having slow spin relaxation and quick decay times, which additionally contribute to ac susceptibility. The initial conditions and measurement protocol decide the magnitude and sign of this dynamical term contributing to ac susceptibility. It has been suggested that such kind of out-of-equilibrium properties emerge by the cumulative effect of geometric frustration, disorder, quantum fluctuations, and the cooperative nature of spin dynamics of these materials.

## I. INTRODUCTION

Armed with an in-depth understanding of the exciting properties of strongly topological materials, new design concepts are being developed to emulate alternate computing devices.[1–4] Exotic properties in topological materials emerge from the complex interplay between reduced dimensionality, fundamental symmetries, electron-electron interactions, relativistic spin-orbit interactions, quantum confinement, quantum coherence, quantum fluctuations, and topology of wavefunctions.[5,6] Geometrically frustrated $Dy_2Ti_2O_7$ (DTO) and $Ho_2Ti_2O_7$ (HTO) pyrochlore oxides are one of them and are known for their low-temperature exotic behavior. In these materials, exotic behavior emerges from the cumulative effect of the lattice architecture of Dy/Ho sites, strong crystal field anisotropy, and magnetic dipolar interaction.[7–12] These factors enable to show various phenomena ranging from- crystal field anisotropy and local spin structure-induced dielectric relaxations[13–17], spin ice state [7,11], and emergent magnetic monopole [9,10,18] in different temperatures regimes. Furthermore, DTO and HTO possess quantum-tunneling-dominated spin relaxation regimes up to 13 K and 30 K, respectively[19–23]. In this temperature regime, spin freezing temperature follows $(x-x_c)^{1/2}$ type variation with the magnetic field (non-thermal variables) as observed for quantum phase transitions.[24–26]

Recently, Samarakoon et al. performed a low-temperature ultrasensitive magnetic noise experiment on DTO, revealing that spin dynamics show cooperative and memory effect behavior.[27] Savary et al.[28] investigated the role of disorder in HTO and proposed the formation of disorder-induced spin liquid-like states governed by quantum correlations. In a previous study, we also observed anomalous thermal hysteresis in magnetic ac susceptibility measurements, which strongly depend on driving frequency.[29] Furthermore, the magnetization study performed by Sakakibara et al.[30] showed the presence of two successive field-induced hystereses below 0.36 K in DTO. They concluded that observed hysteresis is associated with spin ice to kagomé ice and kagomé ice to 3in-1out or 1in-3out spin crossover. A similar result was also found by Yukio et al.[31,32] for $Yb_2Ti_2O_7$ (YTO) quantum spin ice in a temperature-dependent magnetization and neutron diffraction study. These observations indicate that magnetic interaction/field-induced spin crossovers are liquid-gas-type transitions in these materials.

These findings signify the complexity of the magnetic behavior of DTO and HTO, where the interplay of local spin structures, disorder, quantum fluctuations, and the cooperative nature of spin dynamics cumulatively decides the macroscopic properties. Herein, we take a step to uncover this complexity through rigorous magnetic and dielectric investigations. Our findings reveal the crucial role of thermal and non-thermal variables in the out-of-equilibrium state, which dictates the macroscopic properties of these materials.

## II. EXPERIMENTAL DETAILS

High-quality polycrystalline $Dy_2Ti_2O_7$ (DTO) and $Ho_2Ti_2O_7$ (HTO) samples have been used for the magnetic and dielectric measurements. Samples synthesis, phase purity, and their structural details have already been reported in Ref[14,33,34]. Magnetic measurements are performed using Magnetic Properties Measurement System (MPMS-3)® (Quantum Design, Inc.) USA. The samples used were powder and mounted in a brass holder for the measurements. The temperature sweep rate was kept fixed at @1.5 K/min while varying the other non-thermal variables for all the temperature-dependent measurements. Low-temperature magnetodielectric measurements of DTO were performed using an Agilent E4980A LCR meter interfaced with Physical Properties Measurement System (PPMS-3)® (Evercool Quantum Design, Inc.) USA.

## III. RESULTS

Fig. 1 (a &b) shows the temperature-dependent real part of ac susceptibility ($\chi'(T)$) measured for DTO in field-cooled cooling (FCC) and field-cooled warming (FCW) mode for varying ac variables. Fig. 1 (a) shows the $\chi'(T)$ in FCC and FCW mode measured at 1.5 Oe ac amplitude ($H_{ac}$) for different frequencies (f). FCC and FCW of $\chi'(T)$ show a weak frequency-dependent thermal hysteresis ($\Delta\chi'$) below 4 K. Herewith, below 2.4 K, a frequency-dependent crossover in $\Delta\chi'$ is observed for the measured frequency range. For clarity, we denote this hysteresis as an anomalous hysteresis. Below and above the crossover temperature, $\Delta\chi'$ changes its sign, i.e., if we defined $\Delta\chi' = \chi'_{FCC} - \chi'_{FCW}$ then at crossover temperature ($T_{Cross}$), $\Delta\chi'$ become zero and below $T_{Cross}$, it turns out to be positive from negative. On increasing frequency, a peculiar increase in $T_{Cross}$ up to 298.3 Hz occurs and becomes saturated above this frequency. On the other hand, in $H_{ac}$-dependent variation, $T_{Cross}$ is independent with increasing $H_{ac}$ up to 5 Oe and decreases abruptly above 5 Oe.

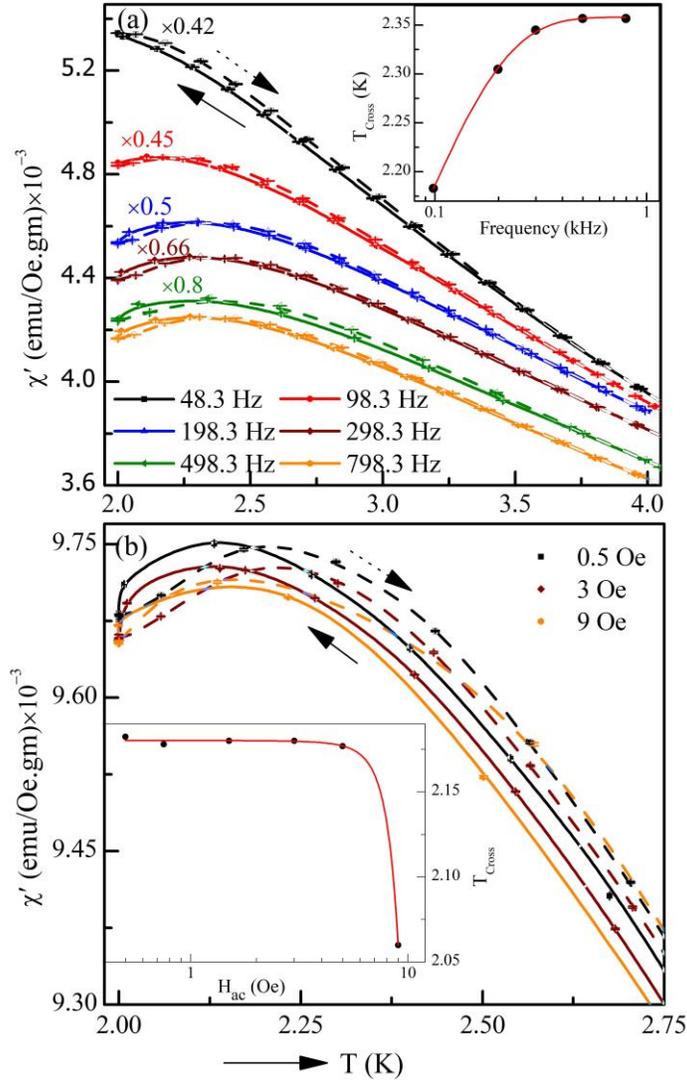

FIG. 1: For $Dy_2Ti_2O_7$ (a) Temperature-dependent real part of ac susceptibility ($\chi'$) measured in FCC (solid symbol and line (smoothen data)) and FCW (open symbol and dashed line (smoothen data)) mode at 1.5 Oe ac amplitude for different frequencies. Inset shows the variation in the crossover temperature ($T_{Cross}$) with frequency. (b) Temperature-dependent $\chi'$ measured in FCC and FCW mode at 98.3 Hz for different ac amplitudes. The inset shows the variation in the $T_{Cross}$ with ac amplitude.

Fig. 1(b) shows the FCC and FCW of $\chi'(T)$ plot for selected 0.5, 3, and 9 Oe ac amplitudes, whereas the complete $\chi'(T)$ plot of FCC and FCW for measured $H_{ac}$ is shown as Fig. S1 (supplementary file). In this measurement, we do not observe any significant change in the magnitude of $\Delta\chi'(T)$, like the frequency. It is noteworthy that we find the crossover in $\chi'(T)$ only. In M(T), thermal hysteresis emerges below 5 K and increases with lowering temperature without any crossover, as shown in Fig. S2(a) (supplementary file). We observe a similar M(T) behavior for HTO. Inset of Fig. 1(a & b) shows the f and $H_{ac}$ dependent variation in $T_{Cross}$ and its fit using the exponential decay equation given by-

$$T_{Cross}(f) = A \times exp(-R \times \chi'(f)) \quad (1)$$

$$T_{Cross}(H_{ac}) = A \times exp(-R \times \chi'(H_{ac})) \quad (2)$$

In equations (1) and (2), A and R are fixed variables representing the amplitude and decay constant, respectively. The deduced values of A and R for $T_{Cross}$ vs. f fit are -0.58±0.024 (K) and (1.217±0.004)×10$^{-2}$ (Oe/emu), respectively. Whereas, for $T_{Cross}$ vs. $H_{ac}$ fit, values of A and R are (-3.68±.05)×10$^{-5}$ (K) and (-89.8±0.1)×10$^{-2}$ (Oe/emu), respectively. A similar measurement protocol has been followed for HTO and is shown in Fig. S3 (supplementary file). In $\chi'(T)$ of HTO, we do not observe any crossover for measured f and $H_{ac}$ range, and it showed similar hysteresis behavior as observed in M(T). In the $\chi'(T)$ measurements of both DTO and HTO, the magnitude of $\Delta\chi'(T)$ does not show any significant change with applied f and $H_{ac}$. The absence of crossover in $\chi'(T)$ of HTO is quite noticeable because DTO and HTO structurally and magnetically mimic each other. In both compounds, ground state Ising doublets separated from the first excited state more than 100 K by strong crystal field anisotropy acting along the <111> direction.[7,35] The deduced values of nearest-neighbor magnetic dipolar ($D_{nn}$) and exchange ($J_{nn}$) interactions are 2.35 K and -1.24 K, respectively, for DTO[11] and 2.4 K and 0.5 K, respectively, for HTO.[7,36] Due to the dominance of ferromagnetic dipolar interaction, both DTO and HTO form exotic spin ice states below 4 K. However, due to the Kramer nature of the $Dy^{3+}$ ion, the magnitude of the transverse field responsible for quantum tunneling is smaller in DTO than HTO.[36,37] It leads DTO to possess a slower spin relaxation time (~ms) than HTO (~ns).[19,21,23] The absence of crossover in HTO indicates that spin dynamics play a vital role along with effective dipolar interaction, which dominates in the crossover temperature range in both compounds.

Fig. 2(a-c) shows the χ′(T) measured in FCC and FCW mode for DTO at different measurement protocols. Fig. 2 (a) shows the χ′(T) measured in FCC and FCW mode for DTO at different dc bias magnetic fields. In this measurement, we observe a linear increase in the $T_{cross}$ with $H_{dc}$ up to 0.01 T (inset of Fig. 2 (a)). A further increase in $H_{dc}$, $T_{Cross}$ expands abruptly with a simultaneous increase in the magnitude of Δχ'.

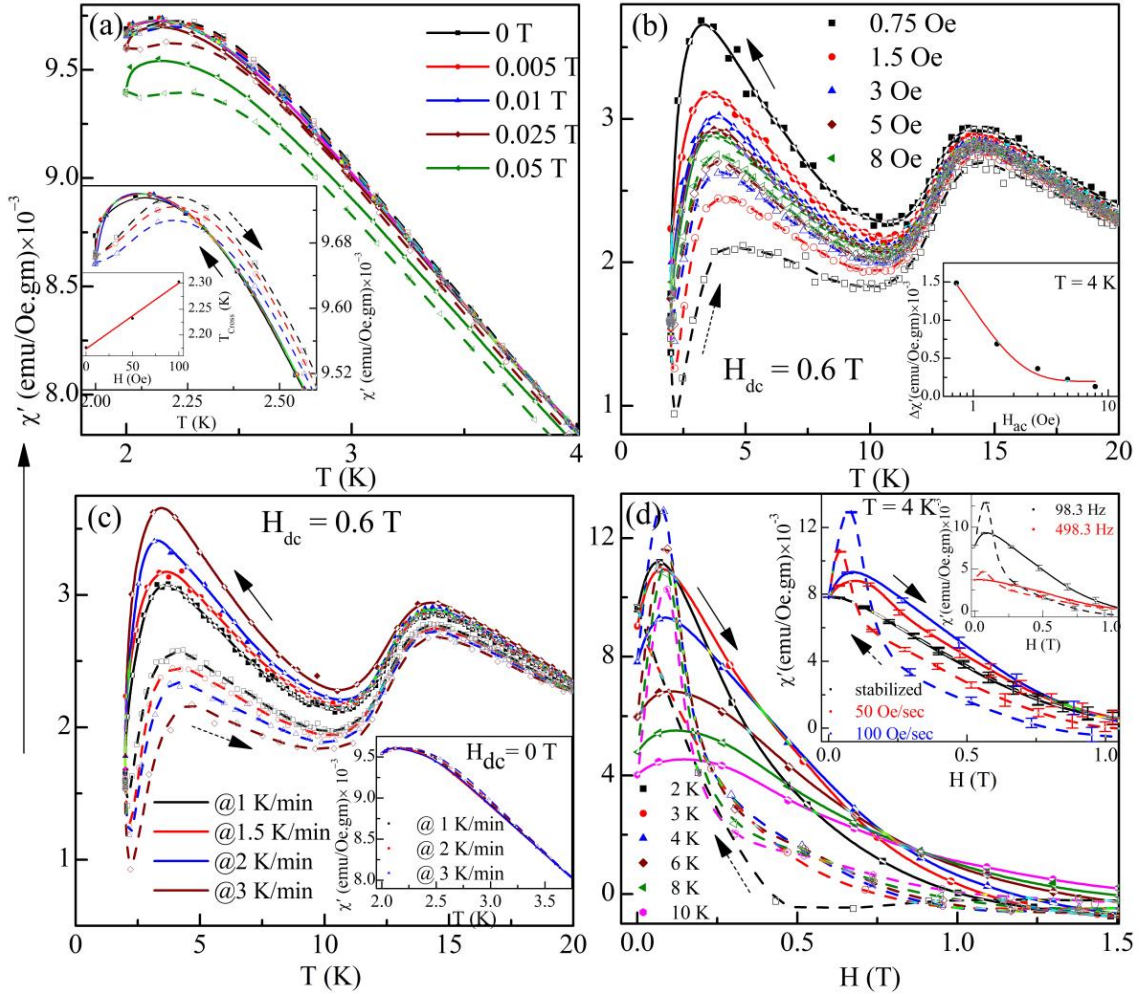

FIG. 2: For $Dy_2Ti_2O_7$, (a) Temperature-dependent real part of ac susceptibility (χ' (T)) measured in FCC (solid symbol and line (smoothen data)) and FCW (open symbol and dashed line (smoothen data)) modes for different dc bias magnetic field at 98.3 Hz and 1.5 Oe ac variables. The inset shows the zoomed view of χ'(T) measured for 0, 0.005, and 0.01 T, $H_{dc}$, and its inset shows the linear variation in crossover temperature ($T_{Cross}$) with $H_{dc}$. (b) χ'(T) measured in FCC and FCW mode at 98.3 Hz frequency for different ac amplitudes in the presence of 0.6 T and 0 T (inset)$H_{dc}$. @ 1.5 K/min temperature sweep rate. (c) χ'(T) measured in FCC and FCW mode at 98.3 Hz and 1.5 Oe ac amplitude for different temperature sweep rates in the presence of 0.6 T and 0 T (inset) $H_{dc}$. (d) Magnetic field-dependent χ' measured in forward 0→1.5 T (solid symbol and line (smoothen data)) and backward 1.5→0 T (open symbol and dashed line (smoothen data)) modes at 98.3 Hz and 1.5 Oe for different temperatures; and for stabilized, 50, and 100 Oe/sec sweep rates (inset) at 4 K. The second inset of Fig. (d) shows the χ'($H_{dc}$) measured for 98.3 Hz and 498.3 Hz at 4K.

To investigate the effect of ac variables on the observed anomalous thermal hysteresis, we measured the $\chi(T)$ at 0.6 T optimal dc bias field at different $H_{ac}$ (Fig. 2(b)). At 0.6 T biased $H_{dc}$, $\Delta\chi'$ has maximum magnitude.[29] In this measurement, anomalous $\Delta\chi'(T)$ shows a strong dependency on $H_{ac}$ and follows exponential decaying behavior with increasing $H_{ac}$ (inset of Fig. 2(b)). A similar observation has been found for HTO, as shown in Fig. S4 (supplementary file). This observation is well corroborated with $\chi'(T)$ measured at 3 Oe for different frequencies (Fig. S 5) at 0.75 T biased $H_{dc}$ for DTO and HTO of our previous work.[29] This feature is exciting because in unbiased $H_{dc}$, $\Delta\chi'$ shows no significant change with ac variables in both compounds. To see the effect of temperature sweep rate on the magnitude of $\Delta\chi'$ for unbiased and biased $H_{dc}$, we performed $\chi(T)$ in FCC and FCW at different temperature sweep rates. In unbiased $H_{dc}$, the magnitude of $\Delta\chi'(T)$ is nearly independent of temperature sweep rate (inset of Fig. 2(c)). Contrary to this, in 0.6 T biased $H_{dc}$, the magnitude of anomalous $\Delta\chi'(T)$ shows linear dependence with temperature sweep rate (Fig. 2(c)). Fig. S6 (supplementary file) shows the $\Delta\chi'(T)$ vs. temperature sweep rate plot and its linear fit. We have plotted the slope vs. temperature plot from the obtained values of the linear fit and shown in the inset of Fig. S6. It has been found that on decreasing temperature slope is linearly up to ~4 K, whereas below 4 K, it decreases. We observe a similar trend for anomalous $\Delta\chi'(T)$ vs. temperature plot.[29] This observation shows the suppression of anomalous $\Delta\chi'(T)$ with temperature.

To understand the role of $H_{dc}$ in the emergence of anomalous hysteresis, we performed the magnetic field-dependent ac susceptibility measurements ($\chi(H)$) on DTO and HTO by sweeping the $H_{dc}$ up to 1.5 T (and cycled back) at a constant temperature. The inset of Fig. 2(d) shows the magnetic field-dependent real part of ac susceptibility ($\chi'(H)$) measured at 98.3 Hz frequency and 1.5 Oe $H_{ac}$ at 4 K for stabilized mode, 50 Oe/sec, and 100 Oe continuous magnetic field sweep rates for DTO. As we can see in the inset of Fig. 2(d), $\chi'(H)$ shows the magnetic hysteresis for 50 Oe/sec and 100 Oe/sec continuous magnetic field sweep mode. Whereas for the stabilized mode, where measurement occurs after stabilization of the magnetic field at each set point, we do not observe any magnetic hysteresis. Furthermore, like the dependency of $\Delta\chi'(T)$ on ac variables (f and $H_{ac}$) and temperature sweep rate, magnetic hysteresis ($\Delta\chi'(H)$) also shows the dependency on f (inset of Fig. 2(d)) and magnetic field sweep rate. Out of these observations, we observe a crossover at ~0.1 T during 1.5 T →0 T magnetic field sweeping in both 50 Oe/sec and 100 Oe/sec sweep rates, below which $\Delta\chi'(H)$ changes its sign similar to the $\Delta\chi'(T)$.

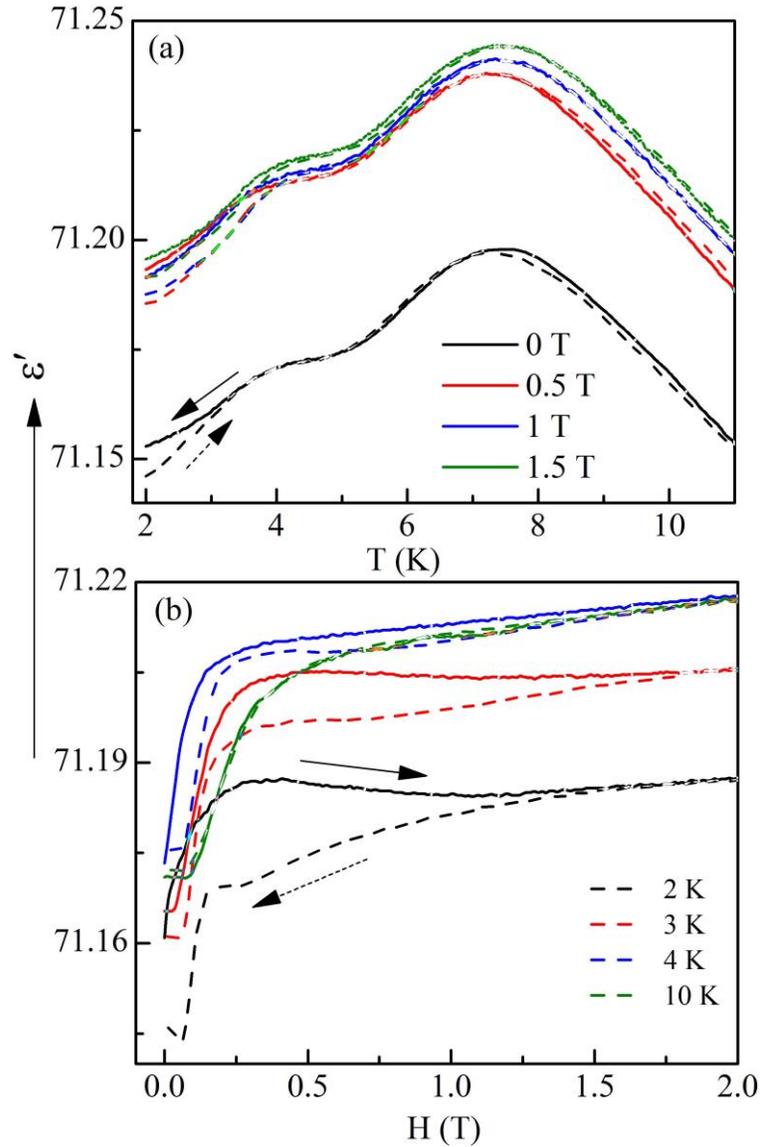

FIG. 3: (a) Temperature-dependent real part of dielectric permittivity ($\varepsilon'$) measured in cooling (solid line) and warming (dashed line) mode @ 2 K/min sweep rate at different magnetic field and (b) magnetic field dependent $\varepsilon'$ measured in forward (0→2 T, solid line) and reverse (2→0 T, dashed line) mode at different temperatures @ 100 Oe/sec sweep rate, for $Dy_2Ti_2O_7$ at 50 kHz frequency.

However, in $\chi'(H)$ measurement performed at different temperatures (Fig. 2(d)), we observed crossover only ≥4 K. A similar observation has been found in HTO as well as shown in Fig. S7 (supplementary file). The plot of the $H_{dc}$ sweep rate and temperature dependence of $\Delta\chi'(H_{dc})$ at different $H_{dc}$ for both compounds is given in the supplementary file as Fig S8 with a detailed discussion.

The observed temperature and magnetic field sweep rate dependence of anomalous hysteresis indicates that the existing phonon-mediated spin-flipping mechanism[15,16,38,39] gets altered by biased $H_{dc}$. To investigate this, we performed the temperature and magnetic field-dependent dielectric permittivity ($\varepsilon$) measurement for DTO. Fig. 3(a) shows the temperature-dependent real part of dielectric permittivity ($\varepsilon'(T)$) measured at different $H_{dc}$. In the $\varepsilon'(T)$ plot (Fig. 3(a)), we observe two successive relaxations at ~7.5 K and 4 K. Furthermore, below 4 K, it shows thermal hysteresis in cooling and warming measurement mode, like M(T) and $\chi'(T, H_{dc}= 0T)$ magnetic measurements. On application of $H_{dc}$, an abrupt increase in the magnitude of $\varepsilon'(T)$ with a slight increase in both relaxation temperatures takes place. On increasing $H_{dc}$, the temperature span of thermal hysteresis increases with a gradual decrease in it′s magnitude. Fig. 3(b) shows the magnetic field-dependent real part of dielectric permittivity ($\varepsilon'(H_{dc})$) measured at different temperatures. In $\varepsilon'(H)$ measurement, a spontaneous increase in the magnitude of $\varepsilon'$ occurs with the magnetic field and becomes saturated above a critical field depending on the temperature. This abrupt increment in $\varepsilon'$ starts at 0.05 T magnetic field and saturates ≤ 0.5 T for measured temperature. Along with this, forward and reverse magnetic field sweeping shows hysteresis in the $\varepsilon'$. The observed hysteresis in $\varepsilon'(H)$ decreases with increasing temperature and vanishes at 10 K. The emergence of thermal and magnetic hysteresis in $\varepsilon'$ below 4 K confirms the strong interconnection of $\varepsilon'$ with local spin structure and spin dynamics.

## IV. DISCUSSION

The above experimental results clearly show the measurement protocol-dependent nature of hysteresis in both DTO and HTO. In M(T) (Fig. S2) and $\chi'(T, H_{dc}=0$ T) (Fig. 1 & Fig. S3), below 4 K, observed thermal hysteresis is independent of temperature sweep rates and ac variables. On application of $H_{dc}$, thermal hysteresis gets suppressed in M(T), whereas in χ′(T), it abruptly expands in the measured temperature range (Fig. 2(a)). Intriguingly, anomalous hysteresis emerges in ac χ only in both compounds. Snyder et al.[19] investigated the effect of $H_{dc}$ on the spin relaxation time (τ) of DTO through frequency-dependent ac χ measurements. They found non-monotonic dependence of τ with the $H_{dc}$ below ~13 K, where τ becomes weak temperature dependent. At the smaller field ($H_{dc}$< 0.5 T), τ(H) decreases with increasing $H_{dc}$ but increases for higher fields and then saturates for the higher $H_{dc}$. The value of the critical field up to which τ(H) decreases shows temperature dependence. At 6 K, the value of the critical field is ~0.65 T whereas, at 12 K, it becomes ~0.35 T. They concluded that applied $H_{dc}$ suppresses the thermal fluctuation and enhances the quantum effects acting on cooperative spins. In the present results, the magnitude of anomalous Δχ′ increases up to ~0.6 T and decreases for the higher $H_{dc}$.[29] The temperature and $H_{dc}$-dependent behavior of anomalous Δχ′ (Fig. 2) corroborates with τ behavior. Further, the spontaneous increment in ε′ with $H_{dc}$ (Fig. 3) also indicates the suppression in the spin-phonon coupling by $H_{dc}$. Scharffe et al. [40] found a decrease in the thermal conductivity coefficient with magnetic field in both DTO and HTO, which supports this argument. Now the question is, what does bias $H_{dc}$? In these materials, Ising spins are situated non-collinearly at the vertex of corner-sharing tetrahedra and pointing towards the center. Due to the non-colinear spin structure and polycrystalline sample, splitting energy (ΔE) of the Ising doublet state has multiple values in the presence of $H_{dc}$. As a result, applied $H_{dc}$ produces random quenched disorder sites having larger spin relaxation times in the Ising spin matrix.[41] These quench disorder sites block the dynamical pathway of the ideal system and provide a limited path for spin-flipping. Due to this restriction, the system takes more time to achieve thermal equilibrium from the out-of-equilibrium state. The suppression in the spin-phonon coupling, which might occur at the quench sites only, supports the increased out-of-equilibrium time duration by $H_{dc}$.

Interestingly, in all $H_{dc}$ biased χ′(T) measurements, we do not observe any significant change in the freezing temperatures in FCC and FCW modes. It indicates that in both measurement modes, the resonant frequency of the cooperative spins at freezing points is the same. Further, the exponential increment in the magnitude of anomalous Δχ′ with decreasing frequency) indicates another time region (large spin relaxation time) in out-of-equilibrium. The temperature and magnetic field sweep rates dependent measurements show the decay of Δχ′(T, $H_{dc}$) in a short time t. It means that when we move the system in the out-of-equilibrium state in the presence of $H_{dc}$, quench disorder sites develop a dynamic spin-correlation of larger τ and exist for a short time t. This field-induced dynamic correlation additionally contributes to the χ′ in out-of-equilibrium. Thus the χ′(T, $H_{dc}$) can be represented as a sum of stationary part ($χ′_{ST}$) and non-stationary part ($χ′_{NST}$):[42]

$$χ′(ω, t) = χ′_{ST} + χ′_{NST} \qquad (3)$$

The $χ′_{NST}$ contribution in χ′ is decided by the value of τ with respect to some characteristic time ($τ_{char}$) at constant temperature and $H_{dc}$. For τ ≥ $τ_{char}$, dynamics is non-stationary, and $χ′_{NST}$ becomes non-zero, whereas for τ ≤ $τ_{char}$, dynamics is stationary and $χ′_{NST}$ becomes zero. In temperature and magnetic field-dependent ac χ measurement, we observe that the value of $χ′_{NST}$ is $\frac{Δχ′}{2}$. The addition in $χ′_{ST}$ takes place, i.e., $χ′_{NST}$ is finite and positive when $k_BT$ decreases (FCC in χ′(T)) or ΔE increases (increasing $H_{dc}$ in χ′(H)). Whereas a subtraction in $χ′_{ST}$ takes place, i.e., $χ′_{NST}$ is finite and negative when $k_BT$ increases (FCW in χ′(T)) or ΔE decreases (decreasing $H_{dc}$ in χ′(H)). The change in sign, dependency on $k_BT$, and ac variables indicate that quantum fluctuation governs the emergent dynamic spin-correlations.[43–45] Where relaxations take place through quantum channels, and properties depend on the initial conditions.[46–48]

In DTO, the emergence of crossover in Δχ′(T, $H_{dc}$=0 T) below 2.5 K indicates that, similar to the biased $H_{dc}$, effective dipolar interaction works similarly. However, unlike $H_{dc}$, magnetic dipolar-induced out-of-equilibrium spin-correlations should differ because each spin feels equal potentials raised by magnetic dipolar interaction. This scenario becomes clearer in magnetic field-dependent χ′ measurements (Fig. 2d), where we do not observe any crossover below 4 K in both DTO and HTO. It leads us to conclude that effective dipolar interaction also develops out-of-equilibrium spin-correlations of short decay time. Due to this reason, out-of-equilibrium properties measured in the absence of $H_{dc}$ were observed at the mK scale, as reported in previous studies.[27,49,50] The absence of crossover in HTO for measured frequency and $H_{ac}$ suggests the speedy decay time of the non-stationary part, most probably due to its faster spin dynamics even at the mK scale. In an equilibrium state, these systems become state independent and behave classically, and show quasi-liquid-gas type transition during Ising paramagnetic to spin ice crossover.[30–32] Due to this cruciality, these systems show complex behavior, even at low temperatures.[30,51]

## V. CONCLUSIONS

Our results demonstrate the distinct behavior of cooperative spins dynamics in the equilibrium and out-of-equilibrium states in DTO and HTO. In the equilibrium state, static and dynamic magnetic properties of both DTO and HTO are weakly dependent on external thermal and non-thermal variables. Both compounds show liquid-gas type transition during interaction/magnetic field-induced local spin state crossover when measured in the equilibrium state. Contrary to this, in the out-of-equilibrium state, we observe a magnetic field-induced thermomagnetic hysteresis found in ac $\chi(T, H)$ only and shows a strong dependency on thermal and non-thermal variables. In dielectric measurements, the abrupt increment in the dielectric permittivity with the applied magnetic field (<0.5 T) indicates that the magnetic field increases the out-of-equilibrium time spam of cooperative spins by weakening the spin-phonon coupling strength. It has been suggested that applied $H_{dc}$ produced quenched disorder sites in the non-collinear Ising spin metrics and developed dynamic spin-correlations with slow relaxation governed by the quantum fluctuations. These dynamic spin-correlations also contribute to $\chi(T, H)$ when the system falls into the out-of-equilibrium state. Based on measurement protocol, the non-stationary part can have positive and negative values and becomes zero in equilibrium or deteriorating conditions. The temperature/magnetic field sweep rate dependence, generally observed for classical systems, shows the short decay time of the out-of-equilibrium state with a complex intertwining of quantum-classical behavior of cooperative spin dynamics in these materials. The observed memory effect in the out-of-equilibrium and its dependency on external stimuli set a stage for understanding the other topologically frustrated quantum materials for future applications.

*SUPPLEMENTARY MATERIAL*

See the supplemental material for the temperature-dependent magnetization measured at different magnetic dc bias magnetic fields, temperature, and magnetic field-dependent ac susceptibility measured for HTO at different ac amplitudes and frequencies with a detailed discussion.

*ACKNOWLEDGMENT*

PY acknowledges the CSIR-HRDG, India, for providing funds for this project, and the authors acknowledge CIF, IIT (BHU) for collecting data in the Magnetic Properties Measurement System (MPMS).

*DATA AVAILABILITY*

The data that support the findings of this study are available from the corresponding author upon reasonable request.

*AUTHOR DECLARATIONS*

**Competing Interest**

The authors declare that they have no known competing financial interests or personal relationships that could have appeared to influence the work reported in this paper.